\documentclass[11pt,a4paper]{article}
\pdfoutput=1
\usepackage{jheppub}

\usepackage{amsfonts,mathtools}
\usepackage{slashed}
\usepackage{xcolor}
\usepackage{comment}
\hypersetup{
    colorlinks,
    linkcolor={red!70!black},
    citecolor={blue!70!black},
    urlcolor={blue!80!black}
}

\newcommand\bra[1]{\left\langle {#1} \right| }
\newcommand\ket[1]{\left| {#1} \right\rangle}
\newcommand{\order}[1]{\mathcal{O}({#1})}

\newcommand\nbar{{\bar{n}}}
\newcommand\ubar{\bar{u}}

\def\figref#1{{figure\ \ref{#1}}}

\def\ncol{{$n$-collinear}}
\def\nbcol{{$\nb$-collinear}}
\def\etab{{\bar\eta}}
\def\nb{{\bar n}}
\def\ni{{n_i}}
\def\nib{{\nb_i}}

\def\chibarn#1#2{{\bar{\chi}_{n#2}^{#1}}}
\def\chinbar#1#2{{\chi_{\nb#2}^{#1}}}
\def\chin#1#2{{\chi_{n#2}^{#1}}}

\def\M{{\cal M}}
\def\B{{\cal B}}

\def\BnH#1#2{{\B^{#1}_{n#2}}}
\def\BnbH#1#2{{\B^{#1}_{\nb#2}}}

\def\J{{\mathcal{J}}}
\def\Jnnb#1#2{{J_{n\nb#2}^{#1}}}

\def\JnO#1{{J_{n0}^{#1}}}
\def\JnOb#1{{J_{n\bar 0}^{#1}}}
\def\Jlo{O_2^{(0)}}
\def\Jnlo#1{O_2^{(1#1)}}
\def\Jnnlo#1{O_2^{(2#1)}}

\def\eperp{\xi}

\def\bigO#1{{\mathcal{O}(#1)}}
\def\oas{{\bigO{\alpha_s}}}

\def\fvM#1#2{{{#1}^{\!\!\!\!#2}}}

\def\sp#1#2{{\,#1\cdot#2\,}}

\def\vect#1{{\mathbf{#1}}}

\def\vbar{\bar{v}}
\def\ubar{\bar{u}}

\def\alimit{{\ensuremath{q^n\bar q^{\nb}g^n} limit}}
\def\blimit{{\ensuremath{q^n\bar q^{\nb}g^{\nb}} limit}}
\def\climit{{\ensuremath{q^n\bar q^{n}g^{\nb}} limit}}

\title{Renormalization of Dijet Operators at Order $\mathbf{1/Q^2}$ in Soft-Collinear Effective Theory}

\author{Raymond Goerke}
\emailAdd{rgoerke@physics.utoronto.ca}
\affiliation{Department of Physics, University of Toronto, 
Toronto, Ontario, Canada M5S 1A7}

\author{and Matthew Inglis-Whalen}
\emailAdd{minglis@physics.utoronto.ca}

\abstract{

We make progress towards resummation of power-suppressed logarithms in dijet event shapes such as thrust, which have the potential to improve high-precision fits for the value of the strong coupling constant.  Using a newly developed formalism for Soft-Collinear Effective Theory (SCET), we identify and compute the anomalous dimensions of all the operators that contribute to event shapes at order $1/Q^2$. These anomalous dimensions are necessary to resum power-suppressed logarithms in dijet event shape distributions, although an additional matching step and running of observable-dependent soft functions will be necessary to complete the resummation. In contrast to standard SCET, the new formalism does not make reference to modes or $\lambda$-scaling. Since the formalism does not distinguish between collinear and ultrasoft degrees of freedom at the matching scale, fewer subleading operators are required when compared to recent similar work. We demonstrate how the overlap subtraction prescription extends to these subleading operators. 
}

\begin{document}

\maketitle
\flushbottom

\section{Introduction}

Event shapes such as thrust, broadening, and the C-parameter are strong probes of the predictive power of perturbative quantum chromodynamics. Thrust \cite{Becher:2008cf,Abbate:2010xh,Abbate:2012jh} and the C-parameter \cite{Hoang:2015hka} have been used to precisely determine $\alpha_s$ from $e^+ e^-$ collision data with the help of Soft-Collinear Effective Theory
 (SCET) \cite{Bauer:2000ew,Bauer:2000yr,Bauer:2001ct,Bauer:2001yt} 
 . The value of $\alpha_s$ determined by these event shape measurements deviates from the world average \cite{Bethke:2013}, and a better understanding of QCD power corrections could help understand this discrepancy. 

For a concrete example, we consider the event shape thrust. Thrust is defined as
\begin{equation} \begin{aligned} \label{eq:thrust_defn}
\tau=1-T=1-\max_{\vect{\hat t}} \sum_{i\in X} \bigg| \frac{\vect{\hat t}\cdot \vect{p}_i }{Q} \bigg|  \ ,
\end{aligned} \end{equation}
where $\vect{\hat t}$ is the unit vector that maximizes the weighted sum over all final state momenta $X$. The value of $\tau$ ranges between $1/2$ for spherically symmetric distribution of momenta in the final state to $0$ for exactly collinear back-to-back jets. A corresponding observable for thrust is the cumulative thrust distribution for $e^+e^-\rightarrow \text{jets}$. Normalizing to the Born cross-section $\sigma_0$ this cumulative distribution $R(\tau)$ is given by
%
%
\begin{equation} \begin{aligned} \label{eq:rate_defn}
R(\tau)=\frac{1}{\sigma_0}\frac{1}{2Q^2} \sum_i L^i_{\mu\nu} \sum_{X}  \int\!d^4x \bra{0} \J_{i}^{\mu\,\dagger}(x) \ket{X} \theta(\tau-\hat{\tau}(X)) \bra{X} \J_{i}^\nu(0) \ket{0},
\end{aligned} \end{equation}
where $\hat\tau(X)$ is the function that computes eq. \eqref{eq:thrust_defn} for each final state $X$. The dependence on the leptonic current has been absorbed in $L_i^{\mu\nu}$, and the current $\J_{i}^\mu=e^{-iq\cdot x}\sum_{f,c}\bar{\psi}_f^c\, \Gamma_i^\mu \psi_f^c$ is either the vector ($\Gamma_V^\mu =\gamma^\mu$) or axial ($\Gamma_A^\mu =\gamma^\mu\gamma_5$) QCD quark current. In the limit $\tau\ll1$ the cumulative thrust distribution can be computed in a perturbative expansion in both $\alpha_s$ and $\tau$, and takes the value \cite{Kramer:1986mc}
\begin{equation} \begin{aligned} \label{eq:rate_result1}
R(\tau)=1+\frac{\alpha_s C_F}{2\pi}( -2\log^2\tau-3\log\tau+2\zeta_2-1 )+\frac{\alpha_s C_F}{2\pi}\tau(2\log\tau-4)+\order{\alpha_s^2,\tau^2}.
\end{aligned} \end{equation} 

For sufficiently small values of $\tau$, the quantity $\alpha_s \log\tau$ becomes large and 
the validity of the asymptotic expansion in fixed-order perturbation theory breaks down. Effective field theories and renormalization group techniques can been used to resum infinite subsets of these logarithmic terms, which improves the validity of the approximation. Summing the infinite subset of Sudakov (double) logarithms, starting with the term proportional to $\alpha_s \ln^2\tau$, is called the leading logarithm (LL) approximation, with NLL describing the summation of the terms starting with $\alpha_s \ln\tau$, and so on. SCET has enjoyed a great deal of success in summing these logarithmic terms up to $\mathrm{N^3LL}^\prime$ in the thrust distribution\cite{Schwartz:2007ib,Becher:2008cf,Chien:2010kc}.

However, the terms suppressed by powers of $\tau$ still limit the theoretical uncertainty in the regime where $\tau$ is small but still large enough that perturbation theory is valid, i.e. $Q\tau \gg \Lambda_{\text{QCD}}$ (also  known as the ``tail'' region of the distribution). Power corrections have been included in thrust calculations using direct and effective field theory methods \cite{Becher:2008cf,Abbate:2010xh,Abbate:2012jh,Moult:2016fqy,Boughezal:2016zws}, however these have been at fixed-order in perturbation theory, which computes only the leading terms in the infinite subset.

In this paper we make progress towards the goal of summing the whole series of leading logarithms suppressed by $\tau$ in the cumulative thrust distribution by computing the anomalous dimensions of all the necessary scattering operators in SCET that contribute to the $\order{\alpha_s\tau}$ cumulative distribution.
We use a new formalism for SCET developed in \cite{Goerke:2017ioi}. The remaining ingredient required to complete the summation is to match onto and renormalize the subleading soft functions. The soft functions in this formalism will correspond exactly to the soft functions discussed in standard SCET for event shapes, which encode the effect of low energy radiation on the event shape. They are therefore observable-dependent, in contrast to the operators we consider in this paper, which apply to any event shape for $e^+e^-\rightarrow \text{dijets}$. A detailed analysis of soft function matching and renormalization in this formalism is currently a work in progress and so we will not discuss it further here\footnote{Definitions of subleading soft functions were given in \cite{Freedman:2013vya}, although the formalism of SCET used there is different from the one used here.}. 

\section{Formalism Review} \label{sec:refresher}

The formalism for SCET developed in \cite{Goerke:2017ioi} expanded on the work of \cite{Freedman:2011kj}, in which SCET was constructed as an effective field theory of decoupled copies of QCD interacting with each other via Wilson lines. This idea was further explicated in \cite{Feige:2013zla,Feige:2014wja} to study factorization in QCD. In \cite{Goerke:2017ioi}, the formalism was modified to remove the ultrasoft sector from dijet operators in the effective theory below the matching scale, while modifying the standard ``zero-bin'' prescription to make the theory consistent. In this section we briefly review the notation and formalism used in this framework for SCET and demonstrate the matching calculation onto the subleading operators.

\subsection{Notation}

Throughout this paper we will use the usual lightcone coordinates:
 \begin{equation} \begin{aligned} \label{eq:light_cone_decomposition}
p^\mu = n\cdot p \frac{\nbar^\mu}{2}+ \nbar\cdot p \frac{n^\mu}{2} + p_\perp^\mu = p^+  \frac{\nbar^\mu}{2}+ p^- \frac{n^\mu}{2} + p_\perp^\mu  \, ,
\end{aligned} \end{equation}
where $n^\mu=(1,\vect{n})$, $\nbar^\mu=(1,-\vect{n})$ and $n\cdot\nbar=2$; we will also use the shorthand $p^{\mu} = (p^+,p^-,\vect{p_{\perp}})$. In standard SCET specific ``$\lambda$-scaling'' is assigned to each light-cone component depending on which sector the particle is in. In contrast, in this formalism there is no need to compare the relative scaling of collinear modes to soft, ultrasoft, or other modes, and so defining a $\lambda$-counting for different components of momenta will not be necessary. When matching onto the effective theory we will consider the limits of QCD in which $p^+$ or $p^-$ are much less than the matching scale $Q$, considering all such perturbations to be of the same order. The power counting of subleading operators in this formalism is then determined entirely by their dimension, as will be made evident below. In the dijet limit, thrust scales like the hemispherical mass-squared $\tau \sim M_H^2/Q^2$ \cite{Hornig:2011iu}, so to calculate the cumulative thrust distribution 
up to $\order{\tau}$ it is necessary to determine the subleading operators up to a suppression of $1/Q^2$. 

We define the following gauge-invariant operator building-blocks which we will use to construct subleading operators, using notation familiar from existing SCET literature:
\begin{equation}\begin{aligned}
\chi_{\ni}(x) &= W^{\dagger}_{\ni}(x)P_{\ni}\psi_{\ni}(x),\\
\mathcal{B}_{\ni}^{\mu_1\cdots\mu_N}(x) &= W_{\ni}^{\dagger}(x)iD_{n_i}^{\mu_1}(x)\cdots iD_{n_i}^{\mu_N}(x)W_{\ni}(x),\\
\end{aligned}\end{equation}
where $P_{\ni} = \slashed{n}_i\slashed{\nb}_i/4$, and $\ni$ are the directions of each jet. For dijets, we always work in a reference frame where $\vect{n}_1 = \vect{n}$ and $\vect{n}_2 = \vect{\nb}$, such that $\vect{\bar{n}}_1 = \vect{n_2}$ and $\vect{\bar{n}}_2 = \vect{n}_1$. The Wilson lines are defined in the usual way,
 \begin{equation} \begin{aligned} \label{eq:wilson_lines}
W_\ni(x)&=\mathcal{ \overline{P} }\exp \bigg[ \! -\! i g\int_{0}^{\infty}   \! \! \! \! ds \, \nib \cdot A_\ni^a(x+\nib s) T^a e^{-\epsilon s} \bigg] \ ,
\end{aligned} \end{equation}
where $\mathcal{\overline{P}}$ denotes antipath ordering and $\epsilon$ is the Feynman pole prescription.  The only distinction between these objects and their equivalents in standard SCET literature is that in 
the present formalism, fields are regular QCD fields with quantum numbers labeling their corresponding sector. 

While the operators in the effective theory depend on the choice of a direction $\vect{n}$, they are in fact invariant under boosts along that direction; i.e. the form of the operators does not depend on which reference frame one uses to define $n^{\mu} = (1,0,0,1)$, provided $\vect{n}$ points along the same axis. Thus, following \cite{Goerke:2017ioi}, we define new vectors $\eta$ and $\bar\eta$,\footnote{Note that these 
differ by a sign from the definitions in \cite{Goerke:2017ioi}, since in our case $q$ is timelike rather than spacelike}
\begin{equation} \label{etas} \begin{aligned}
\eta^{\mu} &= \sqrt{\frac{q\cdot \nb}{q\cdot n}}\; n^{\mu}\\
\end{aligned}
,\quad
\begin{aligned}
\bar\eta^{\mu} &= \sqrt{\frac{q\cdot n}{q\cdot \nb}}\; \nb^{\mu} \ ,\\
\end{aligned} \end{equation}
where $q^{\mu} = (Q/\alpha,Q\alpha,\vect{0})$ 
is the momentum transfer of the process, and in the case of $e^+e^-\rightarrow X$ it is the momentum of the virtual electroweak boson. The parameter $\alpha$ defines the relative boost from the frame in which $n^{\mu} = (1,0,0,1)$. The four-vectors $\eta$ and $\bar\eta$ have been defined so that $p\cdot\eta$ and $p\cdot\bar\eta$ don't depend on $\alpha$ for any $p$, and are therefore useful for making this boost symmetry of the effective theory manifest.

For brevity, we define some shorthand notation to denote the displacement of fields from the interaction vertex in position space, which will be necessary for renormalization:
\begin{equation} \begin{aligned}
\mathcal{B}_{\ni}^{\mu_1\cdots\mu_N}(x,t) &= \mathcal{B}_\ni^{\mu_1\cdots\mu_N}(x+\bar\eta_i t/Q)\\
\chi_{\ni}(x,t) &= \chi_n(x+\bar\eta_i t/Q).
\end{aligned}
\end{equation}
Here $t$ is dimensionless parameter that displaces the fields from the vertex at $x$ along the $\vect{n}_i$ direction. 

Following the lead of \cite{Moult:2015aoa,Kolodrubetz:2016,Feige:2017} we will also find it useful to build subleading operators using a set of building blocks that project out states with definite helicity.  We find that this both simplifies the structure of subleading operators but also allows us to take advantage of the compact form of matrix elements of massless QCD between states with definite helicity. 

Using the standard basis for transverse polarization vectors,
\begin{equation}
\eperp^{\mu}_\pm = \frac{1}{\sqrt{2}}(0,1,\mp i, 0),
\end{equation}
we define the following combinations of quark-antiquark fields,
\begin{equation}\begin{aligned}
\Jnnb{ij}{\pm}(x,t_1,t_2) &= \chibarn{i}{\pm}(x,t_1)\slashed{\eperp}_{\mp}\chinbar{j}{\pm}(x,t_2)\\
\JnO{ij}(x,t_1,t_2) &= \chibarn{i}{+}(x,t_1)\slashed{\etab}\chin{j}{+}(x,t_2)\\
\JnOb{ij}(x,t_1,t_2) &= \chibarn{i}{-}(x,t_1)\slashed{\etab}\chin{j}{-}(x,t_2) \ ,
\end{aligned}\end{equation}
where $\chi_{n_i\pm}(x) = P_{\pm}\chi_{n_i} =\frac{(1\pm\gamma^5)}{2}\chi_{n_i}(x)$ (these correspond to helicity projections for massless quarks). Here and in the following equation, superscripts $i$ and $j$ are fundamental color indices. We will occasionally drop the second and third arguments denoting the shifts when they are not necessary, i.e. $\Jnnb{ij}{\pm}(x)\equiv\Jnnb{ij}{\pm}(x,0,0)$. We also define helicity projections of the gluon fields:
\begin{equation}
\mathcal{B}^{ij}_{n_ih_1\cdots h_N}(x,t) = \eperp_{h_1\mu_1}\cdots\eperp_{h_N\mu_N}\mathcal{B}_{n_i}^{ij\mu_1\cdots \mu_N}(x,t) \ ,
\end{equation}
where $h_i\in\pm$ are helicity labels, and $\mu_i$ are Lorentz indices.

We would like to finish this section by noting that the power counting of an operator in this formalism is determined entirely by the total mass dimension of its constituent fields. 
By way of example, each field $\chi_{\ni}(x,t)$ contributes 3/2 to the mass dimension of any operator in which it appears, while each insertion of a covariant or partial derivative contributes 1 to the mass dimension. 
In this paper the leading order operator has a mass dimension of 3, so an operator with a mass dimension of $3+n$ is said to be suppressed by $n$ powers of $1/Q$ relative to the 
leading order operator.


\subsection{Matching} \label{sec:matching}

We match the QCD current onto a series of subleading operators organized in an expansion in 
inverse powers of $Q$, the energy of the hard interaction:
\begin{equation} \begin{aligned} \label{eq:SSLO_expansion}
\J^\mu(x) = e^{-iq\cdot x}\Bigg[ C_2^{(0)} O_2^{(0)}(x) &+\frac{1}{Q}\sum_i \int\!dt \, C_2^{(1i)}(t) O_2^{(1i)}(x,t) \\
& +\frac{1}{Q^2}\sum_i \int\!dt \,  C_2^{(2i)}(t) O_2^{(2i)}(x,t) + \mathcal{O}\bigg({\frac{1}{Q^3}}\bigg) \Bigg] \ ,
\end{aligned} \end{equation}
where, as above, $q^{\mu}$ is the momentum transfer of the process. 

The leading order operator in eq. \eqref{eq:SSLO_expansion} is the usual leading order dijet operator. Using the building blocks defined in the previous section, it takes the form
\begin{equation}\label{newway}
O_2^{(0)}(x)=\left(-\eperp^{\mu}_{+}J_{n\nb+}^{ii}(x)-\eperp^{\mu}_{-}J_{n\nb-}^{ii}(x)\right)  
\end{equation}
which has matching coefficient \cite{Manohar:2003vb,Bauer:2003di}
%
%
\begin{equation} \begin{aligned} \label{eq:C2LO}
C_2^{(0)}(\mu)=1+\frac{\alpha_s C_F}{4\pi}\bigg( -\log^2{\frac{-Q^2-i0^+}{\mu^2}} + 3\log{\frac{-Q^2-i0^+}{\mu^2}} +\zeta_2-8 \bigg) \ .
\end{aligned} \end{equation}


In this section we demonstrate tree-level matching from QCD onto SCET currents up to order $1/Q^2$. In \cite{Goerke:2017ioi}, details of the matching calculation for $\Jlo$, $\Jnlo{\perp}$ and $\Jnlo{a}$ were presented using this formalism in the context of deep inelastic scattering. 
The details of the matching procedure for dijets are very similar, but for completeness we will include them here.

Following \cite{Goerke:2017ioi}, we take advantage of the simplified form of matrix elements in massless QCD when the helicities of the external states are specified. It is especially useful to use the spinor-helicity formalism for these calculations, and we follow all of the conventions that can be found in the appendix of \cite{Goerke:2017ioi}. We first match onto a general quark-antiquark final state, denoting
\begin{equation}
\M_{q\pm }\equiv  \langle p_1\mp p_2\pm| \J^\mu |0\rangle \ ,
\end{equation}
where the quark ($p_1$) and anti-quark ($p_2$) are forced to have opposite helicities by angular momentum conservation. The exact result in the full theory is then given by
\begin{equation}\begin{aligned}
\M_{q^{\pm}}=&-\sqrt{\sp{p_1}{\eta}} \sqrt{\sp{p_2}{\eta}}\bar{\eta }^{\mu}+\sqrt{\sp{p_1}{\bar{\eta }}} \sqrt{\sp{p_2}{\bar{\eta }}}\eta^{\mu}\\
&-\sqrt{2} e^{i \phi \left(p_2\right)} \sqrt{\sp{p_1}{\bar{\eta }}} \sqrt{\sp{p_2}{\eta}}\xi_{+}^{\mu}+\sqrt{2} e^{-i \phi \left(p_2\right)} \sqrt{\sp{p_1}{\eta}} \sqrt{\sp{p_2}{\bar{\eta }}}\xi_{-}^{\mu} \ .
\end{aligned}\end{equation}

We expand this to leading order in the limit where the quark is collinear to the $\vect{n}$ direction while the antiquark is collinear in the opposite direction, $-\vect{n}$. According to the definitions in the previous section, this limit corresponds to the limit $\frac{\sp{p_1}{\eta}}{Q}\ll1$ and $\frac{\sp{p_2}{\bar\eta}}{Q}\ll1$. Using $\M^{(i)}$ to refer to the $i^{\text{th}}$ order term in this expansion, we have
\begin{equation}
\M^{(0)}_{q^{\pm}} =-\sqrt{2} e^{\pm i \phi \left(p_2\right)}\sqrt{Q} \sqrt{\sp{p_1}{\bar{\eta }}} \xi_{\pm}^{\mu} \ .
\end{equation}
As expected, this is reproduced by the leading order operator $\Jlo$ defined in eq. \eqref{newway}. At next to leading order in this limit, we find
\begin{equation}
\M^{(1)}_{q^{\pm}} =\sqrt{Q} \left(\sqrt{\sp{p_2}{\bar{\eta }}}\eta^{\mu}-\sqrt{\sp{p_1}{\eta}}\bar{\eta }^{\mu}\right) \ ,
\end{equation}
which is reproduced by the operator
\begin{equation}\begin{aligned}\label{Jperp}
\Jnlo{\perp}(x) =& -\bar\eta^{\mu}\left(i\left(\eperp_+\cdot \partial_n\right)J_{n\nb+}(x)+i\left(\eperp_-\cdot \partial_n\right)J_{n\nb-}(x)\right)\\
&-\eta^{\mu}\left(i\left(\eperp_+\cdot \partial_{\nb}\right)J_{n\nb+}(x)+i\left(\eperp_-\cdot \partial_{\nb}\right)J_{n\nb-}(x)\right)
\end{aligned}\end{equation}
where the subscripts on the derivatives $\partial_i$  indicate that the derivative only acts on fields in the $i$-sector; for example,
\begin{equation}
\left(\eperp_\pm\cdot \partial_n\right)J_{n\nb}^{\pm} = (\eperp_{\pm}^\mu\partial_\mu\bar\chi_n^{\pm})\slashed{\eperp}_{\mp}\chi_{\nb}^{\pm} \ .
\end{equation}
%
As was noted in \cite{Goerke:2017ioi}, the $\Jnlo{\perp}$ operator can be absorbed into the leading order operator $\Jlo$ by a small rotation of $\vect{n}$, and therefore reparameterization invariance implies that the matching coefficient and anomalous dimension of this operator will be the same as the leading order operator to all orders in $\alpha_s$.

There are additional subleading operators at this order that only appear with at least one gluon in the final state, and thus we must expand the QCD matrix elements with three-body final states. However, we can take advantage of the fact that matrix elements of the operator $\Jnlo{\perp}$ are proportional to the total perpendicular momentum of a whole sector. By choosing to match onto three body final states with zero perpendicular momentum in each sector we ensure that $\Jnlo{\perp}$ does not contribute, which also serves to simplify the matching procedure.

\begin{figure}\begin{center}
\includegraphics[width=0.6\textwidth]{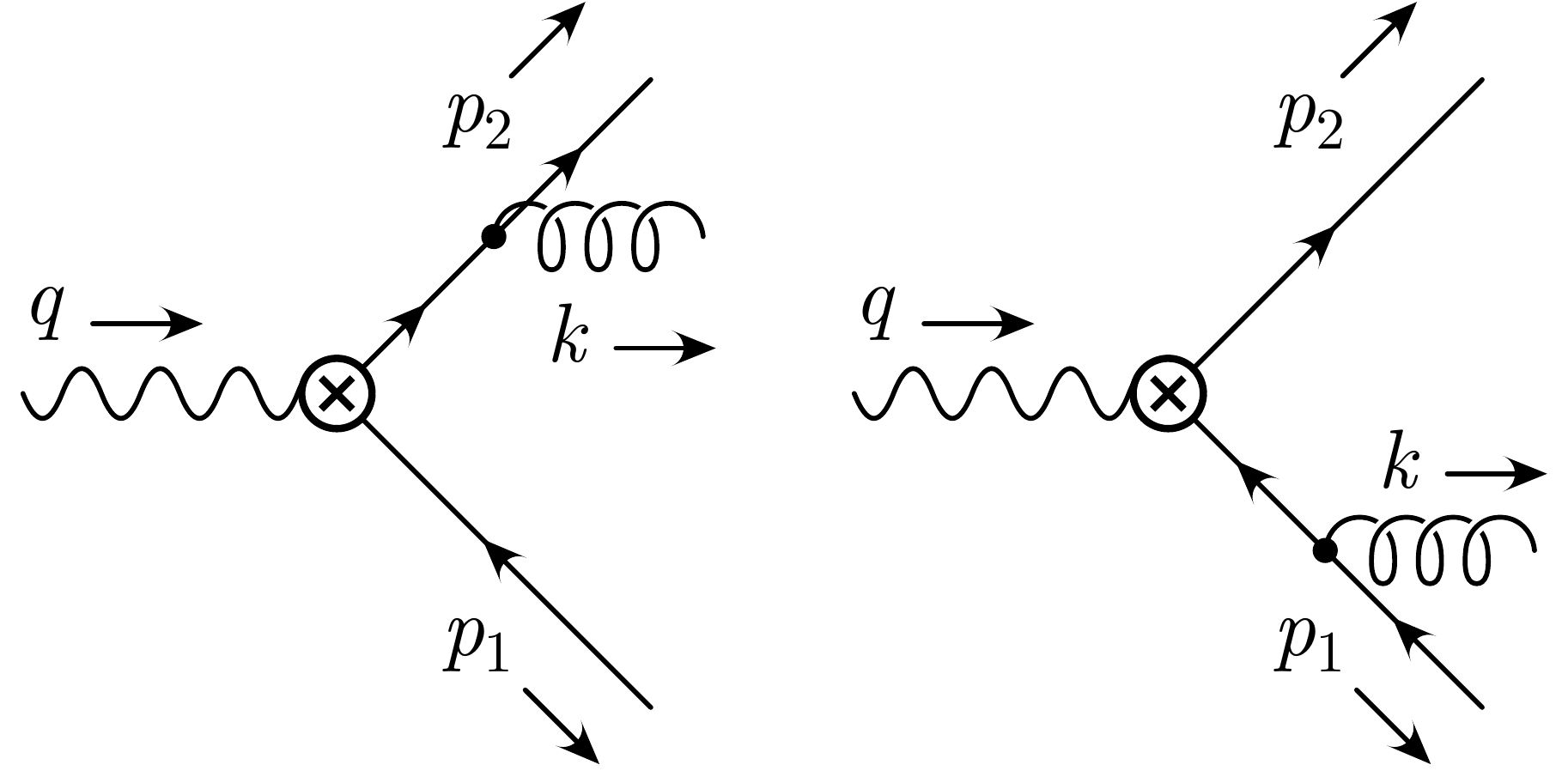}
\caption{QCD graphs contributing to three-body final states.}
\label{fig:j2mu}
\vspace{-0.2in}
\end{center}\end{figure}

The relevant diagrams in QCD for a three-body final state are shown in \figref{fig:j2mu}. For three external particles, there are three ways to combine them into back-to-back sectors. The quark and antiquark can be in different sectors, in which case the gluon can be aligned with either one. Due to the CP symmetry of QCD, these two limits are equivalent, and it will be sufficient to consider the gluon being aligned with the quark. The remaining possibility is that the gluon can be in a sector by itself with the quark and antiquark recoiling together. For brevity, we will refer to these limits by listing the sector of each particle in a superscript, so the first possibility above is the \alimit, which is equivalent to the \blimit, and the remaining case is the \climit. 

We first consider the \alimit\, in which case we arrange the gluon-quark system to have zero perpendicular momentum. Denoting
\begin{equation}
\vspace{-0.05in}
\M_{q\pm g{\pm^\prime}}\equiv  \langle p_1\mp p_2\pm ; k\pm^\prime| \J^\mu |0\rangle,
\vspace{-0.01in}
\end{equation}

where we note that angular momentum conservation ensures the quark and antiquark have opposite helicity, while the helicity of the gluon is independent. We find that the exact result in massless QCD for the diagrams in \figref{fig:j2mu} is
\begin{equation} \begin{aligned}\label{fullqcd}
\M_{g^{\pm}q^{\pm}}=
&-\sqrt{2}gT^a\frac{\sqrt{\sp{p_1}{\bar\eta}}}{\sqrt{\sp{p_2}{\eta}}}\left((\bar{\eta }^{\mu}-\eta^{\mu})-\sqrt2  e^{\mp i \phi (k)} \frac{\sqrt{\sp{p_1}{\eta}}}{\sqrt{\sp{p_1}{\bar{\eta }}}}\xi_{\mp}^{\mu}+\sqrt2  e^{\pm i \phi (k)} \frac{\sqrt{\sp{p_1}{\bar{\eta }}}}{\sqrt{\sp{p_1}{\eta}}}\xi_{\pm}^{\mu}\right)\\
\M_{g^{\pm}q^{\mp}}=
&-2 g e^{\mp i \phi (k)}T^a\frac{ \sqrt{\sp{p_2}{\eta}}}{\sqrt{\sp{p_1}{\eta}}}\xi_{\pm}^{\mu} \ .
\end{aligned} \end{equation}

The leading-order terms of eq. \eqref{fullqcd} in the \alimit\ are already reproduced by the three-body matrix elements of the leading order operator $\Jlo$. Expanding to next-to-leading order in this limit we have
\begin{equation} \begin{aligned}
\M^{(1)}_{g^{\pm}q^{\pm}}=&-\sqrt{2} gT^a \sqrt{\frac{\sp{p_1}{\bar{\eta }}}{Q}} (\bar{\eta }^{\mu}-\eta^{\mu})\\
\M^{(1)}_{g^{\pm}q^{\mp}}=&0,
\end{aligned} \end{equation}
and we find that this is reproduced by the operator
\begin{equation}\begin{aligned}\label{Ja}
\Jnlo{a}(x,t) &= (\bar\eta^{\mu}-\eta^{\mu})\left(\B_{n-}^{ij}(x,t)J_{n\nb+}^{ij}(x)+\B_{n+}^{ij}(x,t)J_{n\nb-}^{ij}(x)\right)\\
\end{aligned}\end{equation}
where $C_2^{(1a)}(t,\mu) = \delta(t) + \oas$. Note that we've included the shift parameter $t$ to displace some fields from the interaction vertex. Despite the fact that matching at tree-level sets $t=0$, a general $t$ will be necessary in order to renormalize this operator, as discussed in the following section.


Now we can match at next-to-next-to-leading order in the \alimit\ onto operators suppressed by factors of $1/Q^2$. Care must be taken in this limit when performing a matching calculation, since momentum conservation relates the three small parameters, i.e. $p_1\cdot\eta$, $p_2\cdot\bar\eta$, and $k\cdot\eta$ are not independent.  The leading order operator $\Jlo$ has matrix elements that can be expanded in $p_1\cdot\eta$ and $k\cdot\eta$, and the higher-order terms must be included consistently to match at $1/Q^2$ (such ambiguities do not appear at $1/Q$). Expanding eq. \eqref{fullqcd} in the \alimit\ to second order and subtracting the corresponding matrix elements of all lower-order effective operators, what is left over at leading order is
\begin{equation}
\begin{aligned}
\M^{(2)}_{g^{\pm}q^{\pm}} =&
2gT^a\left(\frac{ e^{\mp i \phi (k)} \sqrt{\sp{p_1}{\eta}}}{\sqrt{Q}}\xi_{\mp}^{\mu}-\frac{ e^{\pm i \phi (k)} \sqrt{\sp{k}{\eta}} \sqrt{\sp{p_1}{\bar{\eta }}}}{\sqrt{Q} \sqrt{\sp{k}{\bar{\eta }}}}\xi_{\pm}^{\mu}\right)\\
\M^{(2)}_{g^{\pm}q^{\mp}} =& 0 \ .
\end{aligned}
\end{equation}
These terms are reproduced in the effective theory by introducing the operators
\begin{equation}\begin{aligned}
\Jnnlo{a_1}(x,t) &= \left(\fvM{\eperp_+}{\mu}\Jnnb{ij}{+}(x)\BnH{ij}{+-}(x,t)+\fvM{\eperp_-}{\mu}\Jnnb{ij}{-}(x)\BnH{ij}{-+}(x,t)\right)\\
\Jnnlo{a_2}(x,t) &= \left(\fvM{\eperp_-}{\mu}\Jnnb{ij}{+}(x,t,0)\BnH{ij}{--}(x)+\fvM{\eperp_+}{\mu}\Jnnb{ij}{-}(x,t,0)\BnH{ij}{++}(x)\right)
\end{aligned}\end{equation}
where $C_2^{(2a_1)}(t,\mu), = 2i\theta(t) + \oas$ and $C_2^{(2a_2)}(t,\mu) = -2i\theta(t) + \oas$. 

There will also be operators at this order in the \alimit\ formed by acting total perpendicular derivatives on lower order operators, in analogy with the relationship between $\Jnlo{\perp}$ and $\Jlo$. As in that case, the operators with total perpendicular derivatives can always be absorbed into their lower-order counterparts by slight rotation of $\vect{n}$ and must therefore share the same anomalous dimension as their lower-order counterparts. In this paper we choose to focus only on the operators with new anomalous dimensions and so it will be sufficient to match onto states with zero total perpendicular momentum.

The \blimit\ is completely analogous to the \alimit\ and so we won't repeat the details. One can match onto the equivalent operators with \ncol\ gluon fields replaced by \nbcol\ gluon fields and they will have the same matching coefficient and anomalous dimension as the \alimit\ operators.



The \climit, where the gluon is in a sector by itself with the quark and antiquark in the other sector, requires different types of operators.
The exact result in QCD, now considering a configuration where the quark and antiquark have zero total perpendicular momentum, is
\begin{equation}\begin{aligned}\label{climit}
\M_{g^{\pm}q^{\pm}} =&
\sqrt{2}g\frac{\sqrt{\sp{p_1}{\eta}}}{\sqrt{\sp{p_2}{\bar\eta}}}\frac{Q}{\sp{k}{\eta}}e^{\mp i\phi(p_2)}\bigg((\bar{\eta }^{\mu}-\eta^{\mu})  \\
&-\frac{\sqrt{2} e^{\mp i \phi \left(p_2\right)} \sqrt{\sp{p_1}{\eta}}}{\sqrt{\sp{p_1}{\bar{\eta }}}}\xi_{\mp}^{\mu}
+\frac{\sqrt{2} e^{\pm i \phi \left(p_2\right)} \sqrt{\sp{p_1}{\bar{\eta }}}}{\sqrt{\sp{p_1}{\eta}}}\xi_{\pm}^{\mu}\bigg)\\
\M_{g^{\pm}q^{\mp}} =& 
\sqrt{2}g\frac{\sqrt{\sp{p_1}{\eta}}}{\sqrt{\sp{p_2}{\bar\eta}}}\frac{Q}{\sp{k}{\eta}}e^{\pm i\phi(p_2)}\bigg( (\bar{\eta }^{\mu}-\eta^{\mu}) \\
&-\frac{\sqrt{2} e^{\mp i \phi \left(p_2\right)} \sqrt{\sp{p_2}{\bar{\eta }}}}{\sqrt{\sp{p_2}{\eta}}}\xi_{\mp}^{\mu}
+\frac{\sqrt{2} e^{\pm i \phi \left(p_2\right)} \sqrt{\sp{p_2}{\eta}}}{\sqrt{\sp{p_2}{\bar{\eta }}}}\xi_{\pm}^{\mu}\bigg) \ .
\end{aligned}\end{equation}

Expanding eq. \eqref{climit} to leading order in the \climit\ we find
\begin{equation}\begin{aligned}
\M^{(0)}_{g^{\pm}q^{\pm}} =&2 g \sqrt{\frac{\sp{p_1}{\bar{\eta }}}{\sp{p_2}{\bar{\eta }}}}\xi_{\pm}^{\mu} \\
\M^{(0)}_{g^{\pm}q^{\mp}} =&-2g\sqrt{\frac{\sp{p_2}{\bar{\eta }}}{\sp{p_1}{\bar{\eta }}}}\xi_{\pm}^{\mu}
\end{aligned}\end{equation}
which is reproduced by the operator
\begin{equation}\begin{aligned}\label{c-ops}
\Jnlo{c}(x,t) = &\left(\fvM{\eperp_+}{\mu}\JnO{ij}(x,0,t)\BnbH{ij}{+}(x)-\fvM{\eperp_-}{\mu}\JnO{ij}(x,t,0)\BnbH{ij}{-}(x)\right.\\
&\,\,\,\left.\fvM{\eperp_-}{\mu}\JnOb{ij}(x,0,t)\BnbH{ij}{-}(x)-\fvM{\eperp_+}{\mu}\JnOb{ij}(x,t,0)\BnbH{ij}{+}(x)\right)
\end{aligned}\end{equation}
where $C_2^{(1c)}(t,\mu) = 2i\theta(t)$. There is no need to continue the expansion in the \climit\ to higher orders, because operators with this configuration of external states can only interfere with other operators of the same configuration in the calculation of an event shape. Since the operators in eq. \eqref{c-ops} are the leading-order operators in this limit and are already suppressed by $1/Q$ in eq.\eqref{eq:SSLO_expansion}, they are sufficient to consider contributions to the observable at order $1/Q^2$.

This completes the matching procedure required to calculate the fixed order cumulative thrust distribution to $\order{\alpha_s \tau}$ in $e^+ e^-$ scattering. 
Once finished, our 1-loop resummation program is expected to capture the entire leading-logarithmic behaviour of the cumulative thrust distribution.
Pushing the scope of this work to $\order{\alpha_s^2 \tau}$ would require additional tree-level matching with a four-body final state,
though we expect that any contributions resulting from these new operators will be at a lower logarithmic order. 
Further extending the theory
to allow for hadronic initial states would require new gluon-only operators, as shown in \cite{Moult:2015aoa,Kolodrubetz:2016,Feige:2017}, though it should be noted that 
the formalism is those references is different than the one used here.

\section{Renormalization of Subleading Operators} \label{sec:anomdim}

The anomalous dimension of the leading order operator $O_{2}^{(0)}$ has been calculated to three loops \cite{Abbate:2010xh,Lee:2010cga}, and the anomalous dimensions of the subleading $\mathcal{O}(1/Q)$ operators $O_2^{(1a)}$ and $O_2^{(1c)}$ have been calculated to one loop \cite{Freedman:2014uta}. Relevant to the resummation of the $\order{\alpha_s \tau}$ cumulative thrust distribution, there are two operators remaining that have not been renormalized: $O_2^{(2a_1)}$ and $O_2^{(2a_2)}$, and these will be the main results of this paper. In this section, we first review the definition of the overlap subtraction procedure that is used to properly define loop integrals in this formalism, and then we will discuss the definitions and the results of the anomalous dimensions for all the operators we matched onto in the previous sections. 

\subsection{Overlap Subtraction}\label{overlap}

In order to properly define loop integrals in standard SCET, one must introduce the zero-bin subtraction prescription \cite{Manohar:2006nz} or the equivalent, and include both collinear and ultrasoft degrees of freedom in the loops. Formally, the zero-bin removes the overlap of each collinear sector with the ultrasoft sector so as to not double-count degrees of freedom. 

In this formalism
ultrasoft degrees of freedom are not included separately from collinear degrees of freedom in the effective theory, so the subtraction prescription must be modified in order to correctly remove the double-counting, as discussed in \cite{Goerke:2017ioi}. Formally, rather than subtract the overlap of each collinear sector with the ultrasoft sector, one subtracts the overlap between the two collinear sectors. For the calculations we perform here there is little distinction between the two procedures since in each case the zero-bin, overlap, and ultrasoft amplitudes are equal.

We note that for the operators discussed in this paper the overlap subtraction amounts to dividing by the vacuum expectation value of light-like Wilson lines:
\begin{equation}\label{overlap-def}
\frac{\bra{X}O^{(i)}(x)\ket{0}}{\bra{0}\frac{1}{d_\mathbf{R}}\text{tr}\,W_{\nb}^{\mathbf{R}\dagger}(x)W_{n}^{\mathbf{R}}(x)\ket{0}} \ ,
\end{equation}
where each Wilson line is directed along one of the jets and lives in a representation determined by the field content of the operator in the numerator, and where $d_{\mathbf{R}}$ is the dimension of the representation $\mathbf{R}$. For operators in which the quark and antiquark are in different sectors, the Wilson lines are in the fundamental representation and $d_{\mathbf{R}}=N_c$, while for the operators in which the quark and antiquark are in the same sector the Wilson lines are in the adjoint representation and $d_{\mathbf{R}}=N_c^2-1$. 
When eq. \eqref{overlap-def} is expanded in perturbation theory to NLO, it includes a diagram corresponding to the one-loop amplitude of the denominator convoluted with the tree-level amplitude of the numerator, along with 
a minus sign; this formula thus implements the desired subtraction. 

Using eq. \eqref{overlap-def} to define 
the procedure for calculating the one-loop matrix elements of the SCET
operator $O^{(i)}(x)$, we proceed to compute their ultraviolet counterterms and determine their anomalous dimensions.

\subsection{Organization of the Calculation}

To regulate the ultraviolet divergences we use the $\overline{\mathrm{MS}}$ dimensional regularization scheme in $D=4-2\epsilon$ dimensions, and to regulate the infrared divergences we use a gluon mass.  This choice for an infrared regulator provides relatively simple expressions for each loop diagram, with the tradeoff that individual diagrams may contain unregulated divergences \cite{Chiu:2009yx}. 
Despite this drawback, the sum of all diagrams, including the overlap subtraction, is well-defined provided that the integrands are combined before integrating. Since we are computing diagrams with an external gauge boson, we use the background field method \cite{Abbott:1981ke} to make the counterterms gauge-invariant.


We find it most convenient to compute matrix elements in terms of the position-space variable $t$ and then Fourier transform to a momentum space variable $u$ before extracting the counterterms and computing the anomalous dimensions. Formally, the Fourier transformed operators are defined by
\begin{equation} \begin{aligned} \label{eq:fourierO2_t_to_u}
\mathcal{O}_2^{(j)}(x,u) = \int \frac{dt }{2\pi} e^{-iut} \mathcal{O}_2^{(j)}(x,t)
\end{aligned} \end{equation}
and matching coefficients
\begin{equation} \begin{aligned} \label{eq:fourierC2_t_to_u}
C_2^{(j)}(x,u) = \int dt  \, e^{iut}  C_2^{(j)}(x,t) 
\end{aligned} \end{equation}
which together satisfy
\begin{equation} \begin{aligned} \label{eq:parseval}
\int dt\, C_2^{(j)}(x,t) \mathcal{O}_2^{(j)}(x,t) = \int du \, C_2^{(j)}(x,u) \mathcal{O}_2^{(j)}(x,u).
\end{aligned} \end{equation}

Operators of the same $(j)$-label but different value of $u$ mix under renormalization, so that counterterms of $O_2^{(j)}(x,u)$ are non-diagonal in $u$. We write the relation between bare and renormalized operators as
\begin{equation} \begin{aligned} \label{eq:uvCounterterm}
O_{2,bare}^{(j)}(x,u)=\int\!dv\, Z_{2(j)} (u,v) O_{2,ren}^{(j)}(x,v),
\end{aligned} \end{equation}
and we find that operators of different $(j)$-label do not mix under renormalization.

As usual, we note that the bare operators cannot depend on the $\overline{\mathrm{MS}}$ scale $\mu$, so taking the logarithmic derivative of both sides of  eq. \eqref{eq:uvCounterterm} 
and defining the inverse counterterm via the relation
\begin{equation} \begin{aligned} \label{eq:inverseCounterterm}
\int\!dw\, Z_{2(j)}^{-1}(u,w) Z_{2(j)}(w,v) = \delta(u-v)
\end{aligned} \end{equation}
we find the renormalization group equation that governs the running of $O_2^{(j)}(x,u)$
\begin{equation} \begin{aligned} \label{eq:CallanSymanzikO2}
\frac{d}{d\log \mu} O_2^{(j)}(x,u)  = -\int\!dv\,\gamma_2^{(j)}(u,v) O_2^{(j)}(x,v)
\end{aligned} \end{equation}
where the anomalous dimension is 
\begin{equation} \begin{aligned} \label{eq:anomDim}
\gamma_2^{(j)}(u,v) = \int\!dw\,Z_{2(j)}^{-1}(u,w) \frac{d}{d\log\mu} Z_{2(j)}(w,v) .
\end{aligned} \end{equation}
%

Since the combination $\int\!du \, C_2^{(j)}(\mu,u)O_2^{(j)}(\mu,u)$ must be $\mu$-independent, the Wilson coefficient $C_2^{(j)}(u)$ must flow in the opposite manner from its corresponding operator, and in the transposed form
\begin{equation} \begin{aligned} \label{eq:CallanSymanzikC2}
\frac{d}{d\log \mu} C_2^{(j)}(u)  = \int\!dv\,C_2^{(j)}(v)\gamma_2^{(j)}(v,u) \, .
\end{aligned} \end{equation}
Writing the counterterm as a series in $\alpha_s$,
\begin{equation} \begin{aligned} \label{eq:countertermSeries}
Z_{2(j)}(u,v) = \delta(u-v) + \frac{\alpha_s}{2\pi} Z_{2(j)}^{(1)}(u,v) + \mathcal{O}(\alpha_s^2) \ ,
\end{aligned} \end{equation}
the anomalous dimension is then given by:
\begin{equation} \begin{aligned} \label{eq:anomDim2}
\gamma_2^{(j)}(u,v) = \frac{\alpha_s}{\pi} \bigg(\frac{\partial}{\partial \log \mu^2} -  \epsilon  \bigg)Z_{2(j)}^{(1)}(u,v) + \mathcal{O}(\alpha_s^2) \ .
\end{aligned} \end{equation}

\subsection{Results}

\begin{figure}[t]\begin{center}
\includegraphics[scale=1.25]{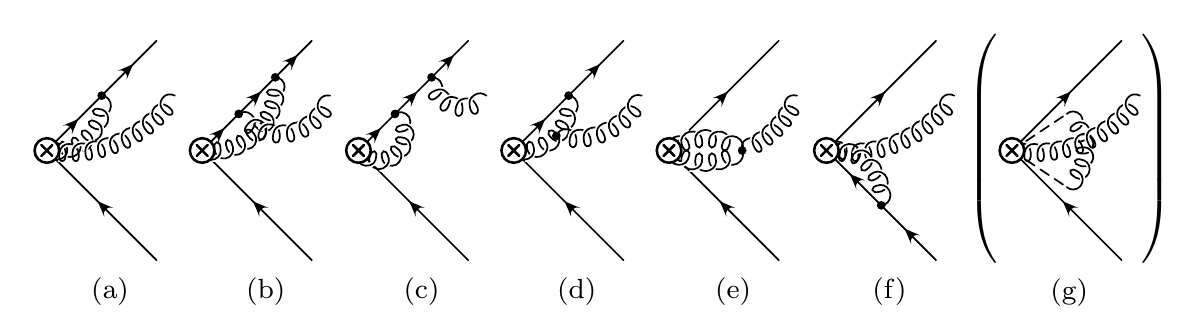}
\caption{The Feynman diagrams for any operator with the $q^n\bar q^{\nb}g^{n}$ configuration. The Feynman rules for the effective vertex are determined by the structure of each operator. Diagram (g) is the overlap amplitude, and must be subtracted.}
\label{fig:A-diagrams}
\vspace{-0.3in}
\end{center}\end{figure}

We first reproduce the results from \cite{Freedman:2014uta}, in which the anomalous dimensions of the operators $O_2^{(1a)}$ and $O_2^{(1c)}$ were computed.

\begin{figure}[h]\begin{center}
\includegraphics[scale=1.25]{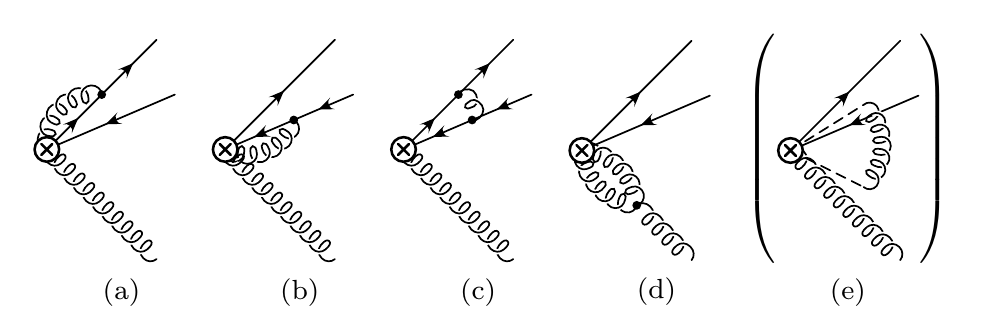}
\caption{The Feynman diagrams for any operator with the $q^n\bar q^{n}g^{\nb}$ configuration. The Feynman rules for the effective vertex are determined by the structure of each operator. Diagram (e) is the overlap amplitude, and must be subtracted.}
\label{fig:C-diagrams}
\vspace{-0.3in}
\end{center}\end{figure}

The relevant diagrams for $O_2^{(1a)}$ operators are shown in \figref{fig:A-diagrams}. To find the counterterms we add together the divergent parts of the diagrams $(a)-(f)$, subtract off the overlap diagram $(g)$, and also include the wavefunction graphs. For $O_2^{(1c)}$ the diagrams are shown in \figref{fig:C-diagrams}; the counterterm is determined by adding diagrams $(a)-(d)$, subtracting the overlap $(e)$, and including the wavefunction graphs. After collecting all the terms and computing the anomalous dimensions according to the notation defined above we find the following results. Note that $u$ corresponds to the fraction of the light-cone momentum $\sp{q}{\nb}$ carried by the particle that was displaced from the vertex in position space; thus, it should be understood that the anomalous dimensions below vanish unless $u\in(0,1)$. For brevity we denote $\bar{u}=1-u$ and $\bar{v}=1-v$.

%
%
\begin{equation} \begin{aligned} \label{eq:anomDim1A}
\gamma_{(1a)}(u,v)&=\frac{\alpha_s\delta(u-v)}{\pi}\left[ C_F \left(\log\frac{-Q^2}{\mu^2}-\frac{3}{2}+\log\bar{v}\right)+\frac{C_A}{2}\left( 1 + \log\frac{v}{\vbar} \right) \right]\\
&+\frac{\alpha_s}{\pi}\left( C_F - \frac{C_A}{2} \right) \ubar \Bigg( \frac{uv}{\ubar \vbar}\theta(1-u-v) + \frac{uv+u+v-1}{uv}\theta(u+v-1)\Bigg) \\
&+\frac{\alpha_s}{\pi}\frac{C_A}{2} \ubar \Bigg( \frac{\vbar-uv}{u\vbar}\theta(u-v)+\frac{\ubar-uv}{v\ubar}\theta(v-u) \\
&\quad\quad\quad\quad-\frac{1}{\ubar \vbar}\bigg[\ubar\frac{\theta(u-v)}{u-v} + \vbar \frac{\theta(v-u)}{v-u}\bigg]_+ \Bigg) \\
\end{aligned}\end{equation}
\begin{equation} \begin{aligned} \label{eq:anomDim1C}
\gamma_{(1c)}(u,v) &=\frac{\alpha_s\delta(u-v)}{\pi}\left[\frac{1}{2}C_F+ C_A\left(\log\frac{-Q^2}{\mu^2}-1+\frac{1}{2}\log v\vbar \right)\right]\\
&-\frac{\alpha_s}{\pi}\left( C_F -\frac{C_A}{2}\right)\frac{1}{v\vbar}\Bigg( v\ubar \theta(u-v)+u\vbar \theta(v-u) \\
&\quad\quad\quad\quad+ \bigg[\ubar v \frac{\theta(u-v)}{u-v} + \vbar u \frac{\theta(v-u)}{v-u}\bigg]_+ \Bigg).
\end{aligned}\end{equation}
We define the symmetric plus-distribution as
\begin{equation} \begin{aligned} \label{eq:plusDefn}
\bigg[ q(u,v) \theta(u-v) + q(v,u) \theta(v-u) \bigg]_+ = \lim_{\beta \rightarrow 0} \frac{d}{d u} &\bigg( \theta(u-v-\beta) \int_{1}^{u} \! \! \! dw\, q(w,v) \\
& + \theta(v-u-\beta) \int_{0}^{u} \! \!  dw\,q(v,w)\bigg)
 \end{aligned} \end{equation}
which satisfies
\begin{equation} \begin{aligned} \label{plusIdentity}
\int_0^1 du \bigg[& q(u,v) \theta(u-v) + q(v,u) \theta(v-u) \bigg]_+f(u)\\
&=\int_0^1 du \bigg( q(u,v) \theta(u-v) + q(v,u) \theta(v-u) \bigg)\left(f(u)-f(v)\right).
 \end{aligned} \end{equation}
Note that we have included fewer operators than in \cite{Freedman:2014uta}, since those authors used a formalism in which ultrasoft degrees of freedom were included in the effective theory below the hard scale. Since we are using a formalism where ultrasoft degrees of freedom are not distinguished from the collinear degrees of freedom below the hard scale, some of the operators defined in that paper have no equivalents in this formalism. We also note that there are some minor errors in the coefficients of the logarithms in the diagonal terms for the equivalent results in \cite{Freedman:2014uta}; we have confirmed that the above results, using the definition of the plus distribution \eqref{eq:plusDefn}, are correct.


We now come to the main result of this paper, in which we present the results for the anomalous dimensions of the $O_2^{(2a_1)}$ and $O_2^{(2a_2)}$ operators, which have been computed for the first time here. The relevant diagrams are also given by \figref{fig:A-diagrams}, as the structure of the graphs will be the same for any operator in which the quark and antiquark are in different sectors. Of course, the Feynman rules to produce a gluon from the vertex is different for each operator. Computing the divergent parts of the graphs, subtracting the overlap graph, and including the wavefunction contributions, we find the anomalous dimensions:
%
%
%
%
%
\begin{equation} \begin{aligned} \label{eq:anomDim2A1}
\gamma_2^{(2a_1)}(u,v) &= \frac{\alpha_s}{\pi}\delta(u-v) \bigg[ C_F \bigg(  \log \frac{-Q^2}{\mu^2} + \log(\vbar) - \frac{3}{2} \bigg)
+ \frac{C_A}{2}\bigg( \log\frac{v}{\vbar} + \frac{5}{2} \bigg)\Bigg] \\
& + \frac{\alpha_s}{\pi} \bigg( C_F - \frac{C_A}{2} \bigg) \frac{1}{v \vbar^2} \bigg( \ubar^2\vbar^2\,\theta(u+v-1)+uv(\ubar\vbar+\ubar+\vbar-1)\theta(1-u-v)\bigg)\\
& - \frac{\alpha_s}{\pi} \frac{C_A}{2}\frac{1}{v\vbar^2}\Bigg( v\ubar^2(1+\vbar)\theta(u-v) +u\vbar^2(1+\ubar)\theta(v-u) 
\\ & \quad\quad\quad\quad
+  \bigg[v\ubar^2 \frac{\theta(u-v)}{u-v} +u\vbar^2 \frac{\theta(v-u)}{v-u}\bigg]_+ \bigg) \, ,
 \end{aligned} \end{equation}
\begin{equation} \begin{aligned} \label{eq:anomDim2A2}
\gamma_2^{(2a_2)}(u,v) &= \frac{\alpha_s}{\pi}\delta(u-v) \bigg[ C_F \bigg(  \log \frac{-Q^2}{\mu^2} + \log(v) - \frac{3}{2} \bigg)
+ \frac{C_A}{2}\bigg( \log\frac{\vbar}{v} + \frac{5}{2} \bigg)\Bigg] \\
& + \frac{\alpha_s}{\pi} \bigg( C_F - \frac{C_A}{2} \bigg) \frac{1}{\vbar v^2} \bigg( \frac{uv}{\ubar\vbar} (\ubar-v)(\vbar-u)\theta(1-u-v) \bigg)\\
&- \frac{\alpha_s}{\pi} \frac{C_A}{2}\frac{1}{\vbar v^2}\Bigg( \frac{v\ubar(\vbar-u)}{\vbar}\theta(u-v) +\frac{u\vbar(\ubar-v)}{\ubar}\theta(v-u) 
 \\ & \quad\quad\quad\quad
+  \bigg[\ubar v^2 \frac{\theta(u-v)}{u-v} +\vbar u^2 \frac{\theta(v-u)}{v-u}\bigg]_+ \Bigg) \, . 
 \end{aligned} \end{equation}


We have used plus-distribution identities to ensure the anomalous dimensions have the form
\begin{equation} \begin{split} \label{eq:symForm}
\gamma_2(u,v)=\delta(u-v) W(v) + f(u,v) S(u,v)\, ,
 \end{split} \end{equation}
 where $W(v)$ is the diagonal part of the anomalous dimension, $f(u,v)$ is analytic in $u$ and $v$, and $S(u,v)$ is symmetric in $u$ and $v$. This property could be important to some readers, since it has previously been exploited to solve the renormalization group equation for the heavy-to-light equivalent of $O_2^{(1a)}$ in terms of Jacobi polynomials \cite{Hill:2004if}.
Extending these methods to the operators $O_2^{(1b)}$, $O_2^{(2a_1)}$, and $O_2^{(2a_2)}$ is outside the scope of this work. 




\section{Conclusion}

We have computed the anomalous dimensions of all operators required to compute subleading corrections to event shapes such as thrust in SCET. We have used a new formalism for SCET that does not make reference to momentum modes or $\lambda$-scaling, and have demonstrated how to match onto a series of higher-dimension operators suppressed by inverse powers of the matching scale $Q$. These anomalous dimensions will be necessary to resum series of subleading logarithms in event shapes, such as those suppressed by powers of $\tau$ in the cumulative thrust distribution, as well as a variety of other event shapes and dijet observables. To complete this program of resummation, an additional matching step onto observable-dependent soft functions will be necessary, and we leave this for future work.


 

\section{Acknowledgements}

We would like to thank Michael Luke for guidance and discussions. This work was supported by the Natural Sciences and Engineering Research Council.

\bibliographystyle{JHEP}
\bibliography{test}

\end{document}